\documentclass[preprint]{epl}

\usepackage[english]{babel}
\usepackage[latin2]{inputenc}
\usepackage[T1]{fontenc}

\usepackage{ae,aecompl}
\usepackage{enumerate}
\usepackage{epsfig}
\usepackage{graphics}
\usepackage{graphicx}
\usepackage{amsmath}
\usepackage{amssymb}
\usepackage{pifont}
\usepackage{wasysym}

\def\ev #1{\left\langle #1 \right\rangle}

\title{Liquidity and the multiscaling properties of the volume traded on the stock market}
\shorttitle{Multiscaling and liquidity}

\author{Zolt\'an Eisler\inst{1}\thanks{Email: \email{eisler@maxwell.phy.bme.hu}} \and J\'anos Kert\'esz\inst{1,2}}
\institute{
  \inst{1} Department of Theoretical Physics, Budapest University of
  Technology and Economics - Budapest, H-1111\\
  \inst{2} Laboratory of Computational Engineering,
  Helsinki University of Technology - Espoo, Finland}

\pacs{89.75.-k}{Complex systems}
\pacs{89.75.Da}{Systems obeying scaling laws}
\pacs{05.40.-a}{Fluctuation phenomena, random processes, noise, and Brownian motion}
\pacs{89.65.Gh}{Economics; econophysics, financial markets, business and management}

\begin{document}

\maketitle

\begin{abstract}
We investigate the correlation properties of transaction data from the New York Stock
Exchange. The trading activity $f_i(t)$ of each stock $i$ displays a crossover
from weaker to stronger correlations at time scales $60-390$ minutes. In both 
regimes, the Hurst exponent $H$ depends logarithmically on the liquidity of the
stock, measured by the mean traded value per minute. All multiscaling
exponents $\tau(q)$ display a similar liquidity dependence, which clearly
indicates the lack of a universal form assumed by other studies. The origin of 
this behavior is both the long memory in the frequency and the size of consecutive transactions.
\end{abstract}

Financial markets are self-adaptive complex systems and their understanding requires
interdisciplinary research, including the application of concepts and tools of
statistical physics. The success of modern statistical physics lies to a large extent in
explaining phenomena from phase transitions to far-from-equilibrium processes, where
the two key concepts have been scaling and universality. When applied to physical
systems, both can rely on a solid foundation: renormalization group theory. But
how reliable can insights based on these principles be, if we move on to social
or economic systems? According to economists, "physicists [simply] suffer from
the belief that there must be universal rules" \cite{culturecrash}.

The aim of present paper is to point out that the assumption of universality
can lead to false conclusions regarding stock market dynamics \cite{gallegatti.etal}. We use
multifractal analysis -- an approach very commonly pursued in econophysics --
to point out that the size of the company, or more appropriately the liquidity
of its stock, affects the observed characteristics of how it is traded on the
market. This dependence is continuous, and therefore it means the absence of universality classes in trading dynamics.

By means of multifractal analysis, we show that: (i) Trading activity records show a crossover from weaker to stronger correlations around the time scale of $1$ trading day. (ii) The strength of correlations above the crossover depends logarithmically on the average trading activity of the stock. (iii) The whole family of $\tau(q)$ multiscaling exponents of trading activity shows a similar variation. (iv) These effects originate from an interplay between the autocorrelations of the frequency and the size of consecutive transactions.

The dataset used in our study was taken from the NYSE TAQ database \cite{taq1993-2003}, and it contains the records of all transactions of the New York Stock Exchange in the years $2000-2002$.

Let us denote the total traded value of a stock $i$ in the time window $[t, t+\Delta t]$ by
$f_i^{\Delta t}(t)$. This is calculated as a sum of the values of all transactions for the given stock during $[t, t+\Delta t]$. If $N_i^{\Delta t}(t)$ denotes the number of trades for the stock in the interval, and the value of the $n$-th trade is $V_i(n)$, then one can write this formally as
\begin{equation}
f_i^{\Delta t}(t) = \sum_{n \atop {t_i(n)\in [t, t+\Delta t]}} V_i(n),
\label{eq:flow}
\end{equation}
where the sum runs for the $N_i^{\Delta t}(t)$ trades in the interval.

Though the returns are known to be only short time correlated, financial
data contain different kinds of long-range correlations, examples range
from volatility to order flow \cite{bouchaud.book, stanley. book, farmer.patience}. Records of traded value are no exception from this \cite{eisler.sizematters, ivanov.itt, ivanov.unpublished},
and are most often characterized by the Hurst exponent, or in general, by multifractal spectra \cite{vicsek.book, dfa}. Multifractal models represent a dynamically developing approach in describing financial processes both in conservative finance and econophysics (for a review, see \cite{bouchaud.multifractal}).

Recent studies \cite{eisler.sizematters, queiros.volume} have shown that the standard deviation, and even higher moments of $f$ exist, thus it is possible, to define the $q$-th order partition function in the following way:
\begin{equation}
\sigma_i^q(\Delta t) = \ev{\left \vert f_i^{\Delta t}(t) - \ev{f_i^{\Delta t}(t)}
\right \vert^q} \propto \Delta t^{\tau_i(q)}, \label{eq:hurst}
\end{equation}
where $\ev{\cdot}$ denotes time averaging. For any fixed stock $i$, the formula defines a $\tau_i(q)$ set of exponents, indexed by $q$, and determined by the slopes of eq. \eqref{eq:hurst} on a log-log plot \footnote{Note that throughout the paper we use $10$-base logarithms.}. These are often written in the form $\tau(q)=qH(q)$, and $H=H(2)$ is called the Hurst exponent, while other $H(q)$'s are the generalized Hurst exponents. This family of exponents is closely related to the correlation properties of the data. If $H=0.5$, the data have no
long range correlations, while for $H > 0.5$ ($H < 0.5$) signals have persistent
(antipersistent) long range correlations. If $H(q) \equiv H$ is independent of $q$, the signal is
self-affine, while nontrivial $q$-dependence gives rise to multiscaling
or multi-affinity.

Here, we present an analysis of the $\sigma_i^q(\Delta t)$ partition functions. We investigated the $2416$ stocks which were continuously listed at NYSE during the years $2000-2002$, and which had an average turnover $\ev{f}$ (mean traded value per minute) of at least $100$ USD/min. This ensures that there are no extended periods where the stock is not traded at all, and thus $f(t)$ is well-defined.

For the calculation of $\sigma_i^q(\Delta t)$ we used Detrended Fluctuation Analysis \cite{dfa}. This method uses piecewise polynomial fits to remove instationarities from the data, and often produces good estimates for $\tau(q)$. We tested the robustness of our estimates to the order of this detrending, and varied the order of the polynomials from $1$ to $5$, but the results did not change significantly.

\begin{figure}[tp]
\centerline{\includegraphics[height=190pt,trim=0 0 0 0]{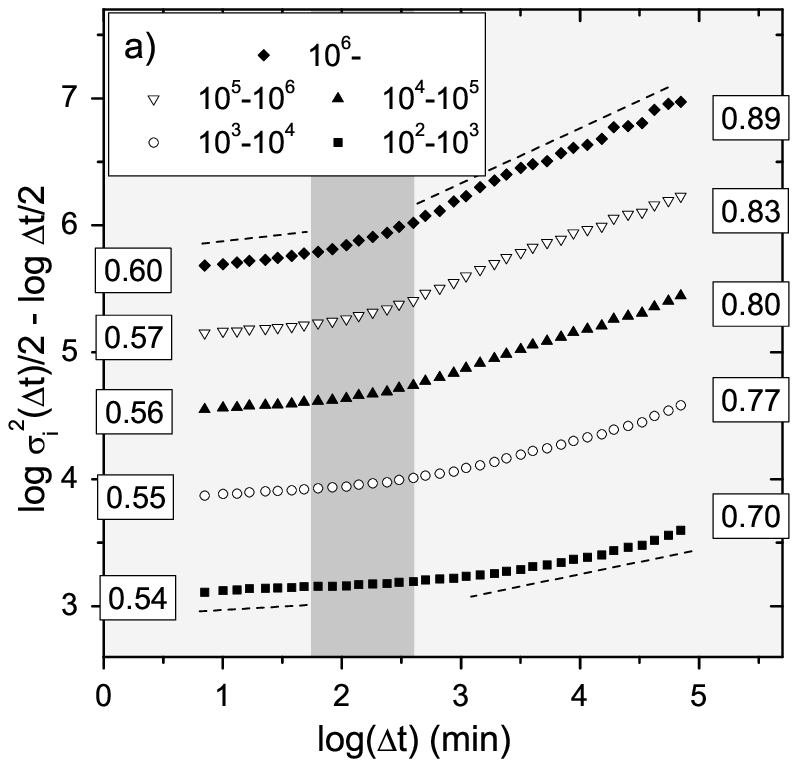}\includegraphics[height=190pt,trim=15 11 20 1]{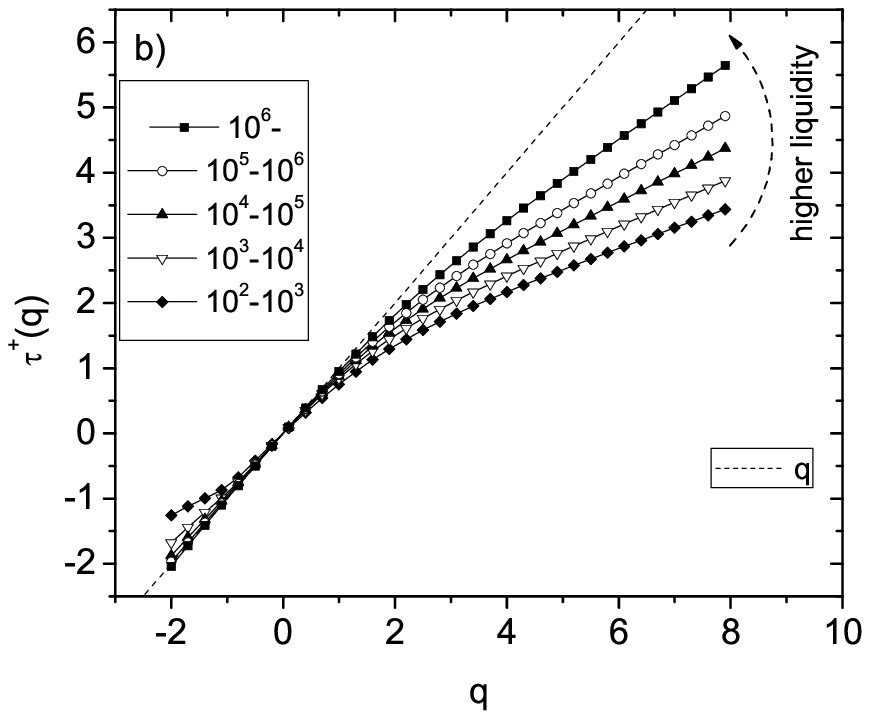}}
\caption{{\bf (a)} The normalized partition function $\frac{1}{2}\log \sigma_i^2(\Delta t)-\frac{1}{2}\log \Delta t$ for the five groups of companies. A horizontal line would mean the absence of autocorrelations in the data. Instead, one observes a crossover phenomenon in the regime $\Delta t = 60-390$ mins. Below the crossover all groups show weakly correlated behavior. Above the crossover (fluctuations on daily and longer scales), the strength of correlations, and thus the slope corresponding to $H^+-0.5$, increases with the liquidity of the stock. \emph{Note}: Fits are for the regimes below $60$ min and above $1700$ min. {\bf (b)} The values of the scaling exponents $\tau^+(q)$, valid for time scales longer than a trading day, for the five groups of companies. Companies with higher average traded value exhibit stronger correlations, and weaker multiscaling than their smaller counterparts. Correspondingly, their $\tau^+(q)$ is greater, and the shape of the curve is closer to the linear relationship $\tau^+(q)=q$.}
\label{fig:Gscavg}
\end{figure}

Then, we divided the stocks into five groups with respect to $\ev{f}$: those with $10^2$ USD/min $\leq \ev{f} < 10^3$ USD/min, those with $10^3$ USD/min $\leq \ev{f} < 10^4$ USD/min, \dots, and finally $10^6$ USD/min $\leq \ev{f}$. Then, we averaged the $\sigma_i^q(\Delta t)$ partition functions within each group \footnote{This averaging procedure decreases the noise present in the data, without affecting our main conclusions. Also note that data were first corrected by the well-known U-shape pattern of daily trading activity (see, e.g., ref. \cite{eisler.non-universality}), calculated independently for each group.}. As an example, the results for $q=2$ are shown in fig. \ref{fig:Gscavg}(a).

One finds that, regardless of group, the $\log \sigma^q(\Delta t)$ versus $\log
\Delta t$ plots are not straight lines. Instead, one observes a crossover
phenomenon \footnote{The fact, that the properties of stock market time series are different on time scales shorter than and longer than $1$ trading day, was pointed out by many sources. The most common examples is are the distribution of returns and the autocorrelations of volatility\cite{bouchaud.book, stanley.book}.}\cite{eisler.sizematters, ivanov.itt}: There are two regimes of $\Delta t$ for which different $\tau(q)$-s
can be identified. For $\Delta t < 60$ min, we are going to use the notation $\tau^-(q)$, while for $\Delta t > 390$ min, $\tau^+(q)$. One can define the related generalized Hurst exponents as $\tau^{\pm} (q)=qH^{\pm}(q)$.
Systematically, $H^+(q)>H^-(q)$, which means that correlations become stronger when window sizes are greater than $390$ min.

Moreover, there is remarkable difference between groups when $\Delta t>390$ min. This means that the correlations present in the day-to-day variations of trading activity systematically depend on $\ev{f}$, as seen from $H^+(q)$ values indicated in fig. \ref{fig:Gscavg}(a). More of this dependence can be understood if one examines the scaling exponents for more powers of $q$. This was done by first evaluating the value of $\tau^+_i(q)$ for the independent stocks, and then averaging that for the elements within each group. The results are shown in fig. \ref{fig:Gscavg}(b). The plot implies, that more liquid stocks (greater $\ev{f}$) display stronger correlations than their less liquid (smaller $\ev{f}$) counterparts, for any order $q>0$. This is realized in a way that the degree of multiscaling decreases, and the scaling exponents tend to
the fully correlated self-affine behavior with the limiting exponents $\tau^+(q)=q$, $H^+(q)=1$.

\begin{figure}[pt]
\centerline{\includegraphics[height=190pt,trim=10 0 0 0]{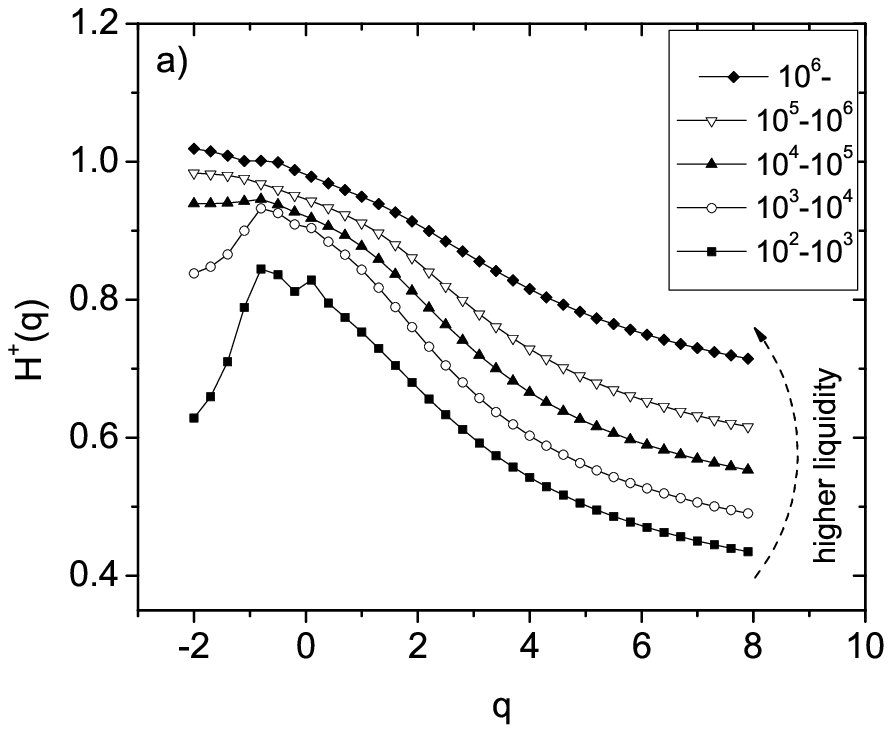}\includegraphics[height=190pt,trim=20 0 0 0]{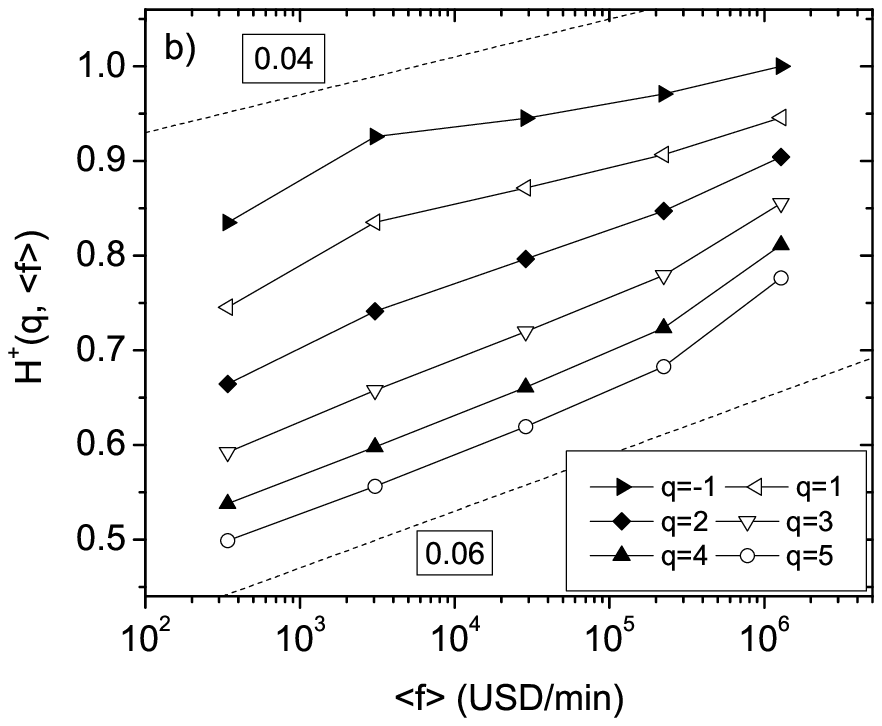}}
\caption{{\bf (a)} The values of the generalized Hurst exponents $H^+(q)$, valid for time scales longer than a trading day, for the five groups of companies. The difference in the strength of correlations, and thus $H^+(q)$, is present for all
powers $q$. This implies, that such a dependence on liquidity is present in both
low and high trading activity periods. {\bf (b)} The values of $H^+(q)$, from top to bottom  $q=-1, 1, 2, 3, 4, 5$. The points represent the average value for one of the five groups of companies. One can see, that $H^+(q)$ changes in an approximately logarithmic fashion with $\ev{f}$. \emph{Note}: Stocks grouped by $\ev{f}$, increasing from bottom to top, ranges given in USD/min.}
\label{fig:Gresavgtauq}
\end{figure}

Fig. \ref{fig:Gresavgtauq}(a) shows the corresponding values
of $H^+(q)$. The difference in the $H^+(q)$'s between the groups is present
throughout the whole range of $q$'s, not only for large $q$'s which are
sensitive to the high trading activity. This indicates that the higher level of
correlations in more liquid stocks cannot be exclusively attributed to periods
of high trading activity. Instead, it is a general phenomenon, that is present
continuously \footnote{One may notice, that there is a strong deviation in the case of stock with low liqudity, and $q<-1$. The origin of this artifact is a finite size effect: The stocks are traded in lots of $100$, and thus they cannot be traded in values less than price$\times 100$. This minimum acts as a cutoff in small fluctuations, to which $q<-1$ moments are very sensitive.}.

Despite the presence of non-universality, and that $\tau(q)$ depends on the liquidity of the stock, there is a clear systematic way \emph{how} this dependence is. In fig. \ref{fig:Gresavgtauq}(b), we plot vertical "cuts" of fig. \ref{fig:Gresavgtauq}(a). These show, that for a fixed value of $q$, $\tau^+(q)$ increases with $\ev{f}$ in an approximately logarithmic way:
\begin{equation}
\tau^+(q; \ev{f})= C(q)+\gamma(q)\log\ev{f},
\label{eq:gammaq}
\end{equation}
where $\gamma(q)\approx 0.04-0.06$.

\begin{figure}[pt]
\centerline{\includegraphics[height=190pt]{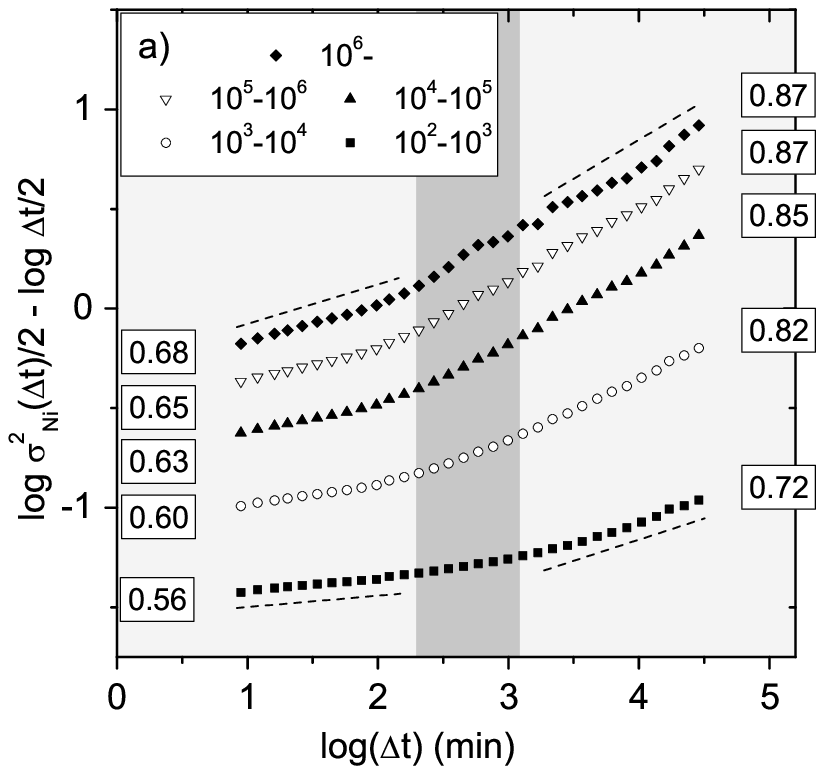}\includegraphics[height=190pt]{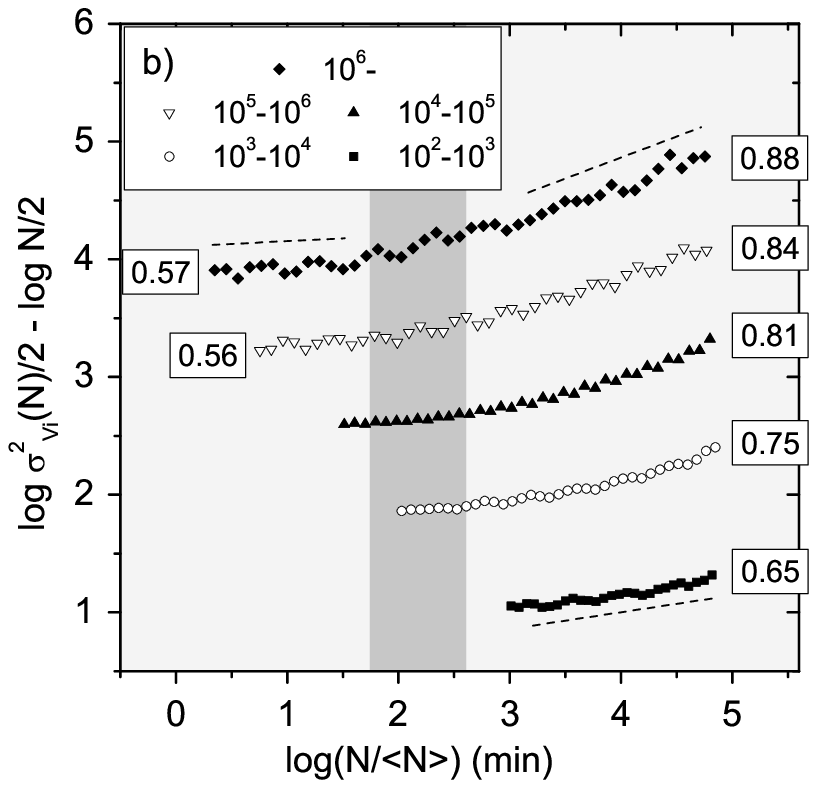}}
\caption{{\bf (a)} The normalized partition function $\frac{1}{2}\log \sigma^2_{Ni}(\Delta t)-\frac{1}{2}\log \Delta t$ for the five groups of companies. A horizontal line would mean the absence of autocorrelations in the data. The crossover regime is for slightly longer times, $\Delta t \approx 160-1200$ min. Above the crossover the strength of correlations in $N$, and thus the slopes corresponding to $H_N^+-0.5$, increase with the liquidity of the stock. increases with the liquidity of the stock. {\bf (b)} Same as (a), but for $\frac{1}{2}\log \sigma^2_{Vi}(N/\ev{N})-\frac{1}{2}\log N$. The darker shaded area corresponds to the crossover regime of $f$ at $N/\ev{N}\approx 60-390$ mins. Small stocks are traded infrequently, therefore they have no data points below the crossover. \emph{Note}: Stocks grouped by $\ev{f}$, increasing from bottom to top, ranges given in USD/min. In both plots, fits are for the regimes below $60$ min and above $1700$ min.}
\label{fig:Gtransposed}
\end{figure}

Our results imply that the trading of assets of companies with very different
size and liquidity cannot be described in a universal manner \footnote{A recent preprint \cite{ivanov.unpublished} shows similar effects with respect to the market where the stocks are traded. More indications of similar behavior can be found in refs.  \cite{eisler.sizematters,bonanno.dynsec}}. There have been
studies pointing out such asset-to-asset variations, and the key role of
liquidity \cite{lubos.liquidity, chordia.liquidity, farmer.whatreally, farmer.powerlaw},
however, they have been consistently overlooked by some econophysics groups.
There is a wide range of studies, that calculate ensemble averages over a large
number of stocks, irrespective of their liquidities. In some cases
universality seems indeed to hold, like for the normalized distribution of returns \cite{lux.paretian, gopi. inversecube}. However, in other cases, as we have just
seen, it is misleading to calculate averages for stocks with a wide range of liquidity as done in, e.g., refs. \cite{gopi.volume, drozdz.average, stanley. commodities, queiros.multifractal}. A "typical" $\tau(q)$ or multifractal spectrum of assets is not meaningful in the presence of this clear, systematic dependence.

What aspect of the trading dynamics is the origin of this non-universality? As eq. \eqref{eq:flow} suggests, the source of fluctuations in $f$ is the fluctuation of $N$ and $V$ (see also ref. \cite{eisler.internal}). Thus, it is very instructive to define the Hurst exponents of these two processes in analogy with eq. \eqref{eq:hurst}. We restrict ourselves to the $q=2$ moment. One can introduce the $H_{Ni}$ Hurst exponent of the time series $N_i^{\Delta t}(t)$ as
\begin{equation}
\sigma^2_{Ni}(\Delta t)=\ev{\left ( N_i^{\Delta t}(t) -\ev{N_i^{\Delta t}}\right )^2} \propto \Delta t^{2H_{Ni}}.
\label{eq:Nhurst}
\end{equation}
This $H_N$ describes the temporal correlations of the number of trades. 


The results for the group averages of $\sigma^2_{Ni}$, and the asymptotically valid exponents $H_{Ni}^\pm$ are shown in fig. \ref{fig:Gtransposed}(a). A comparison with $\sigma^2_i$ [fig. \ref{fig:Gscavg}(a)] shows that both quantities behave similarly: Fluctuations in the number of trades $N$ display crossover and liquidity dependence in the strength of correlations, just like $f$.

The $H_{Vi}$ Hurst exponent of the so-called tick-by-tick data $V_i(n)$ can be defined as
\begin{equation}
\sigma^2_{Vi}(N/\ev{N_i})=\ev{\left (\sum_{n=1}^N V_i(n) -\ev{\sum_{n=1}^N V_i(n)}\right )^2} \propto (N/\ev{N_i})^{2H_{Vi}}.
\label{eq:tickhurst}
\end{equation}
The important point here is that the scaling variable is the $N$ number of consecutive trades. This is divided by the $\ev{N_i}$ mean number of trades per minute. This is crucial, because the
trading frequency of the stocks varies over many orders of magnitudes. Thus $N$
trades corresponds a different time span depending on trading frequency, i.e., on
the stock. The scaling variable $N/\ev{N_i}$ has a dimension of minutes (just
like $\Delta t$), and its fixed value always means the same time
window size, regardless of the stock.

Moreover, when applying eq. \eqref{eq:tickhurst}, there is a natural lower limit in window size: one cannot take less than one trade, and so $N\geq 1$.
Consequently, a group average for $\sigma^2_{Vi}$ is undefined, where the
scaling variable would be $N/\ev{N_i}<1/\ev{N_i}$ for any stock in the group \footnote{We allowed up to $10\%$ of such missing data.}. For more liquid stocks, $\ev{N_i}$ is larger, thus the minimal window size is smaller.

The results are shown in fig. \ref{fig:Gtransposed}(b). $H_V^-$ is only
defined for the two groups, whose stocks are traded at least every $10$ minutes,
and they indicate weak or no liquidity-dependence. $H_V^+$ exists for all groups and
follows the same trend of increasing correlations for greater liquidity.

\begin{table}[tp]
	\centering
		\begin{tabular}{c|ccc|ccc}
		$\ev{f}$ (USD/min) & $H_V^-$ & $H^-$ & $H_N^-$ & $H_V^+$ & $H^+$ & $H^+_N$ \\
		\hline
		$10^6-$ & $0.57$ & $0.60$& $0.68$ & $0.88$ & $0.89$ & $0.87$\\	
		$10^5-10^6$ & $0.56$ & $0.57$& $0.65$ & $0.84$ & $0.83$ & $0.87$\\		
		$10^4-10^5$ & -- & $0.56$& $0.63$ & $0.81$ & $0.80$ & $0.85$\\		
		$10^3-10^4$ & -- & $0.55$& $0.60$ & $0.75$ & $0.77$ & $0.82$\\
		$10^2-10^3$ & -- & $0.54$& $0.56$ & $0.65$ & $0.70$ & $0.72$\\
		\hline
		\end{tabular}
	\caption{Hurst exponents for $f$, $N$ and $V$ for the $5$ groups of stocks. For all groups, every exponent is higher above the crossover than below it. Moreover, above the crossover there are no large differences between $H_V^+$,  $H^+$,  $H_V^+$. From the fits, the errors are estimated to be $\pm 0.03$. \emph{Note}: $H_V^-$ is not defined for the $3$ groups, whose stocks are not traded at least every $10$ minutes.}
	\label{tab:hurst}
\end{table}

The number of transactions in a given time window $[t, t+\Delta t]$ is -- to a good approximation -- independent from the value of the single transactions \footnote{This means, that $N_i^{\Delta t}(t)$ is independent from $f_i^{\Delta t}(t)/N_i^{\Delta t}(t)$. The $R^2$ values of regressions between the logarithms of these two quantities are typically of the order $0.03$ in the data.}. Under this condition, one can show that for any stock $i$:
\begin{eqnarray}
\sigma_i^2(\Delta t)=\sigma^2_{Ni}(\Delta t)\ev{V_i}^2+\sigma_{Vi}^2\ev{\left(N_i^{\Delta t}\right)^{2H_{Vi}}},
\label{eq:proof4}
\end{eqnarray}
where $\ev{V_i}$ is the mean, and $\sigma^2_{Vi}$ is the standard deviation of the value of individual transactions. The origins of the two terms in the formula are the following \cite{eisler.internal}:
\begin{enumerate}
\item The first term describes the effect of fluctuations in the number of transactions. Let us assume, that the size of the transactions is constant, so $V_i(n)=\ev{V_i}$, and $\sigma^2_{Vi}=0$. Then, the second term is zero, and eq. \eqref{eq:proof4} simplifies to $\sigma^2_{Ni}(\Delta t)\ev{V_i}^2$. 
\item The second term describes the effect of fluctuations in the value of individual transactions. If one assumes that the number of transactions is the same in every time window, $N_i^{\Delta t}(t)=\ev{N_i^{\Delta t}}$, then $\sigma^2_{Ni}=0$. The first term becomes zero, and eq. \eqref{eq:proof4} reduces to $\sigma^2_{Vi}\ev{N_i^{\Delta t}}^{2H_{Vi}}$.
\end{enumerate}

Thus the correlations in $f$ originate from the correlations in $N$ and $V$. By definition, the l.h.s. of eq. \eqref{eq:proof4} is proportional to $\Delta t^{2Hi}$. The first term on the r.h.s. is proportional to $\Delta t^{2H_{Ni}}$, while the second term can be estimated to scale as $\Delta t^{2H_{Vi}}$. For large $\Delta t$, the behavior of $\sigma^2$ is dominated by the larger of $H_N$ and $H_V$. 

This is in agreement with the results summarized in table \ref{tab:hurst}. The table also shows that above the crossover there are no major differences between $H$, $H_N$ and $H_V$. This means that neither of the two processes dominates in general.

We have studied the correlation functions \eqref{eq:hurst}, \eqref{eq:Nhurst} and \eqref{eq:tickhurst}
found the following: (i) There exists a crossover in the behavior as a function of the time window at $\Delta t \approx 1$ trading day; (ii) There is non-universal (multi)scaling for large $\Delta t$ with a systematic dependence of the exponents on the liquidity; (iii) Eq. \eqref{eq:proof4} points out the interplay between the fluctuations in the number of trades and the tick-by-tick values, resulting in the observed long term correlations for the activity. While we emphasize the non-universal character
of the exponents, we also mean to underline the systematic trends as a function of the
company size (liquidity). These properties of trading should be addressed by the
future modeling efforts of the stock market.

The authors are grateful to Gy\"orgy Andor for his help with financial data.
JK is member of the Center for Applied Mathematics and Computational Physics, BME. Support by OTKA T049238 is acknowledged.

\end{document}